# Disaggregation for Improved Efficiency in Fog Computing Era

Opeyemi O. Ajibola, Taisir E. H. El-Gorashi, and Jaafar M. H. Elmirghani

*School of Electronic and Electrical Engineering, University of Leeds, LS2 9JT, United Kingdom*

**ABSTRACT**
This paper evaluates the impact of using disaggregated servers in the near-edge of telecom networks (metro central offices, radio cell sites and enterprise branch office which form part of a Fog as a Service system) to minimize the number of fog nodes required in the far-edge of telecom networks. We formulated a mixed integer linear programming (MILP) model to this end. Our results show that replacing traditional servers with disaggregated servers in the near-edge of the telecom network can reduce the number of far-edge fog nodes required by up to 50% if access to near-edge computing resources is not limited by network bottlenecks. This improved efficiency is achieved at the cost of higher average hop count between workload sources and processing locations and marginal increases in overall metro and access networks traffic and power consumption.

## 1. INTRODUCTION

Over the last two decades, the cloud and its associated pay-as-you-go cost model for computational resources has become a household term for end-users, enterprises and governments across the globe. The application of virtualization and other software defined technologies [1], [2] in centralized hyperscale datacentres built using commodity hardware gives cloud infrastructure providers the ability to sustainably offer affordable compute capacity to interested users. To further boost the sustainability of their datacentres, infrastructure providers are known to strategically locate massive datacentre infrastructures near green energy sources [3], [4]. However, present and future proliferation of applications [5] – [9] which require real-time interaction and the expected growth in the uptake of Internet of Things (IoT) paradigm [10], [11] which is estimated to comprise of about 50 billion geographically distributed mobile or static connected things in 2020 [12] have motivated the proposition of a complementary computing tier outside the cloud i.e. Fog computing tier.

Fog computing extends the cloud to the near and far edges of the network via the deployment of compute, storage and networking capabilities in heterogeneous devices and nodes (i.e. fog nodes) such as bespoke servers, edge routers, access points and range of endpoints including connected vehicles, surveillance and cameras just to mention a few. These heterogeneous fog nodes with varying form factors and real estate requirements adopt heterogeneous network infrastructures (both wireless and wired) for connectivity. Collectively, these heterogeneous nodes enable a cloud of things [13] continuum which can support the novel fog as a service (FaaS) business model. One goal of fog computing is to minimize latency for real-time applications (such as game streaming) and mission critical applications by ensuring closeness between the users and their computation locations. Another goal of fog computing is to improve network infrastructure sustainability via the reduction of present and future network infrastructure cost and power consumption that arise from the transmission of massive data generated at network edges by IoT devices and other big data sources to remote centralized cloud for processing [14] – [18]. Adoption of virtualization technologies in fog nodes ensures the enjoyment of some attributes of cloud computing such as virtualization and multitenancy [19] in fog nodes. However, fog nodes are unable to enjoy other benefits obtainable in centralized cloud computing infrastructures such as the ability to strategically locate massive computational resources at little or no opportunity cost and the massive total cost of ownership (TCO) reductions enabled by the economies of scale of commodity hardware used as the basic unit of massive computing clusters.

Furthermore, given a classification of the network edge into two: near edge and far edge, where the near edge of the network comprises of telco's central offices (COs) and radio cell sites (CSs) while the far edge of the network comprises of access points and consumer premise devices, IoT gateways and endpoints such as user devices, sensors and actuators. Relatively, the installed computational capacity in fog nodes at the near edge is less than that available in the cloud. Likewise, the installed computational capacity in fog nodes at the far edge of the network is less than those at the near edge of the network. This is because the size and form factor of real-estate hosting computational capacity drops as one moves from the cloud to the lowest layers of fog computing as shown in Figure 1. (Note that the common presence of small to medium sized private datacentres in enterprise branch offices and research institutes is an exception to this trend). It is expected that the computational resources located in the far edge of the network will suffer from poor utilization due to limited capacity of their network interfaces while the need for multi-hop paths required to access their compute capacity will result in cumulative high access latency that will further discourage their utilization. Hence, to ensure the sustainability and energy efficiency of the access network layer and end devices, minimizing the number of fog nodes at the far edge of the network is desirable to reduce both the TCO and improve the overall energy efficiency of the information and communication technology (ICT) industry [20]. Centralization of fog computing capacity in existing ICT infrastructure locations such as COs, CSs and private computing clusters can aid the realization of this goal, while the interconnecting network topologies can play an important role in optimally placing the fog nodes [21] – [23]. Furthermore, adoption of the server disaggregation concept in such locations can significantly improve their efficiency.

In this paper, we investigate the impact of using COs, CSs and enterprise private computing clusters as fog computing nodes to support real-time applications and elicit enabling communication network criteria for such scenarios. We also study the performance gains of adopting disaggregated servers (DSs) in such locations relative to the use of traditional servers (TSs). These evaluations, studies and investigations are performed via the formulation of a mixed integer linear programming (MILP) model.

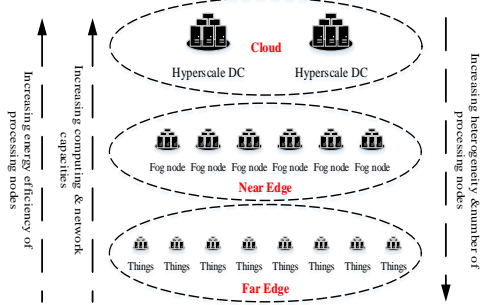
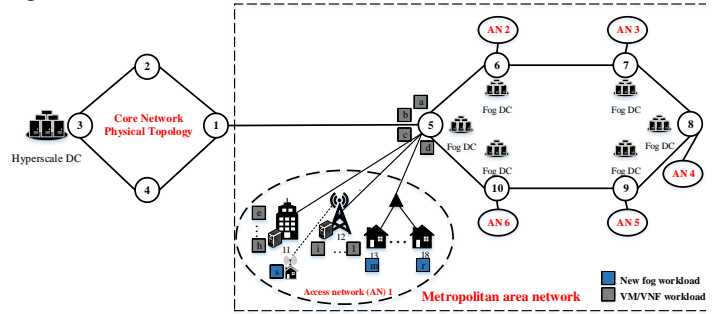

*Figure 1. Characteristics of Cloud of things.*   *Figure 2. Fog as a Service System.*

## 2. FOG AS A SERVICE AND SERVER DISAGGREGATION

Fog as a Service (FaaS) is a new business model which allows small, medium and large service providers to seamlessly deploy and operate computing, storage and control services at different scales [24]. However, such a novel business model requires the appropriate supporting infrastructure, platform, software and evaluation metrics [25]. The proliferation of edge computing nodes such as hyper-converged infrastructure in enterprises is due to performance or regulation requirements. It is also due to the adoption of network function virtualization (NFV) paradigm in telco cell sites and central offices, which promise a pool of computing capacity that can be integrated into FaaS systems to support a wide range of fog computing applications. NFV proposes the virtualization of traditional (physical) network devices as virtual machines (VMs) which run on commodity hardware [26]. By so doing, cloud-like benefits such as increase in consolidation and infrastructure hardware economies of scale can also be enjoyed in the network infrastructure. Furthermore, the transformation of network devices into virtual entities also enables cloud-like agility and innovation in network infrastructure. The adoption of commodity hardware in NFV compliant central offices and cell sites implies that spare computational capacities can be easily used for non-network related services and applications. Likewise, spare computational resources in enterprise branch offices can also be used to support fog applications.

In [27], the authors leveraged on a combination of software defined networking (SDN), NFV and the best features of cloud computing in central offices re-architected as a datacentre with the goal of making the central office an integral part of any future cloud strategy adopted by telco service providers to support new network and non-network applications and services. This concept can be further extended to the other network infrastructure in the access network such as cell sites. The emergence and adoption of the fog-computing paradigm is one of such use cases that can immensely benefit from such evolution of cell site and central office infrastructure given the proximity of these locations to the near and far edges of the network. However, continued adoption of traditional server architectures in such re-architected network infrastructure may inhibit the realization of maximal efficiency as result of known deficiencies of the TS architecture. Hence, it is important to evaluate the impact of server disaggregation in such environments relative to the performance of traditional practises.

Disaggregation proposes the physical or logical separation of TS intrinsic resources into pools of homogeneous resources which can be composed, decomposed and recomposed on-demand via high bandwidth and low latency networks. This concept addresses the limitations associated with TSs such as poor resource modularity and lifecycle management, the need for purpose-built servers in computing clusters, high power consumption and capital expenditure resulting from inefficient resource utilization [28] – [30]. Adoption of these DSs in the fog nodes located in COs, CSs and enterprises BOs can further improve the performance of FaaS business model as this can provide additional fog computing capacity to support real-time applications whilst supporting traditional mission critical application and network functions. In [31], Ericsson proposed a software defined hardware platform characterised by disaggregation and modularity as the solution to reduce overall CAPEX and OPEX in next-generation ICT infrastructure. Other industry leading vendors have also proposed hardware platforms for ICT infrastructure that leverage the benefits of disaggregation [32], [33].

In this paper, we formulate a MILP model that provisions real-time workloads such as online-gaming and IoT intelligent gateway services across distributed fog nodes located in metro COs, enterprise BOs and radio CSs. We give priority to real-time applications in our model because proximity is required between their users and computation locations to ensure optimal performance relative to other fog applications. Our model supports dynamic transitions between TS and DS architectures on-demand. The MILP model optimally places real-time applications with CPU, memory, storage, network requirement within the aforementioned distributed adopted fog nodes while satisfying the set objective function given in Eq. (1). The objective function minimizes the total network power consumption (TNPC) in metro and access networks, the number of fog nodes required at the farthest edge of the network, and the number of active servers (NS) and resource components (NC) required in fog nodes at the near edge of the network. Therefore, the objective is to

$$\textbf{Minimize}: TNPC + \alpha_1 \cdot (\sum_{f \in F} Blocked_f) + \alpha_2 \cdot NC + \alpha_3 \cdot (\sum_{s \in S} NS_s) \quad (1)$$

Where $\alpha_1$ is the cost associated with blocked workload, it is the cost of installing a fog node in the far-edge of the telecom network to support real-time applications; $\alpha_2$ is the TCO associate with the use of each resource component within the server; this cost is active when distributed fog nodes in the FaaS system comprise of DSs; $\alpha_3$ is the TCO cost per active server in distributed fog nodes of the FaaS system. Note that $\alpha_1 \gg \alpha_2 \gg \alpha_3$. These costs implicitly optimize secondary factors of interest such as resource utilization, power consumption and capital expenditure. Hence, to minimize the complexity of the overall MILP model we do not explicitly optimize these secondary factors.

Two different power profiles are adopted for metro and last-mile access network components as shown in Table 1. Access network component such as consumer premise equipment (CPEs) and optical network units (ONUs) are given a static power profile because they are associated with a specific location, hence their power consumption is not shared. On the other hand, optical line terminals (OLT), metro Ethernet access router and metro Ethernet aggregation router are given a load proportional power consumption profile because these components are shared by multiple last-mile access nodes. The value associated with each component's power profile are also given in Table 1. In the MILP model, each real-time workload is associated with a specific access node (workload source) and after successful placement of each workload, a percentage (e.g. 50%) of the workload's uplink data rates is used to commune with the access node that the workload is associated with and to a remote hyper-scale DCs in the internet. This is the first evaluation scenario referred to as "S1" henceforth. We acknowledge that more variations in workload distribution and network traffic patterns can be explored. We leave such evaluations to future work. We scale the workloads' uplink data rate by 100% of its original value to derive the second evaluation scenario referred to as "S2" henceforth. Regular traffic also flows between nodes in the model concurrently with traffic associated with real-time fog applications. We assumed that regular traffic comprises of traditional network traffic and the traffic to and from workloads that do not require real-time processing.

*Table 1. Network component power profile and values.*

| Network Component | Power profiles | Value |
|---|---|---|
| Metro Ethernet customer premise equipment | Static | 75 W [34] |
| Metro Ethernet aggregation router | Load proportional | 0.9 W/Gb [35] |
| Metro Ethernet access router | Load proportional | 0.243 W/Gb |
| PON optical line terminal | Load proportional | 1.75 W/Gb |
| PON optical network unit | Static | 15 W [36] |

*Table 2. Model input parameters.*

| Components/Link | Capacity/Configuration |
|---|---|
| Traditional server (TS) | 12 cores, 64GB RAM and 300GB HDDS |
| CPU resource demand | 3-12 cores |
| RAM resource demand | 20 - 60 GB |
| Storage resource demand | 20 - 120 GB |
| Workload uplink data rate | 1 - 2 Gbps |
| Access node to metro CO links | 10 Gbps |
| Metro CO to metro CO links | 200 Gbps |

## 3. IMPACT OF DISAGGREGATION ON REAL-TIME FOG APPLICATIONS

We consider a small metro network topology comprising of six metro aggregation nodes i.e. COs as illustrated in Figure 2. Connected to each metro CO are an enterprise BO, a radio CS with NFV capable commodity computing hardware and a 1:4 10G EPON last-mile access network. Metro COs, enterprise BOs and radio CSs are fitted with local computing capacity to support traditional VMs or virtual network functions (VNFs) associated with these locations while real-time fog applications are also supported when feasible. We adopt the configuration of the TS given in Table 2 as the basic unit of computation at all fog computation locations in the metro network topology illustrated in Figure 2. This TS is virtually disaggregated over a suitable network fabric to enable the DS architecture in this paper. The number of servers present in each CO, BO and CS are limited to six, six and four servers respectively. Four, four and two node-local workloads (i.e. VMs or VNFs) are associated with each CO, BO and CS respectively while a real-time workload is associated with each CS and residential PON terminal. Each workload has CPU, memory, storage resource demands and uplink data rate which are generated using uniform distribution within the ranges defined in Table 2. Table 2 also gives the capacity of links of the metro/access network topology in Figure 2.

The model can be used to evaluate per component power consumption and utilization in each adopted fog node. However, we limit our focus to the number of blocked real-time workloads, number of active components and/or servers, the total network traffic in the metro and access networks and the resulting total network power consumption associated with them. We do not discuss the power consumption in the core network based on the assumptions that the interactive workloads can only be provisioned in the metro and access network tiers for optimal performance. Hence, these workloads have limited impact on core network power consumption. If an interactive workload cannot be provisioned within the adopted fog nodes in the metro topology in Figure 2, it is blocked. However, blocking in this case implies that a local computing capacity must be installed at the source of the interactive workload to support it, therefore leading to higher TCO, higher power consumption and lower overall energy efficiency.

Whilst the link capacity between COs and all access nodes is 10 Gbps and real-time workload uplink data rate is not scaled, Figure 3 shows that 3 interactive applications are blocked when TSs are deployed in COs, BOs and CSs that have been adopted as fog nodes. These applications are blocked inspite of the presence of 15 idle TSs in BOs and CSs distributed across the metro network topology as shown in Figure 4. Failed access to the computing resources in these idle TSs over the network is responsible for this because the links connecting the BOs and CSs to their adjacent COs lack enough capacity to support the network traffic associated with the blocked workloads. The impact of this phenomenon on TSs in CSs is managed by the model (relative to TSs in BOs) via the strategic placement of interactive workloads requested by users which are directly tethered to the telecom network via wireless links provided by CSs. These workloads are provisioned on the CS which gives them access to the metro network. This ensures that the capacity of links between such CSs and their adjacent COs is conserved to support remote access of computing resources in the local CS. Notwithstanding, 13% of unused servers are in CSs while 87% are in BOs. On the other hand, no real-time application is blocked when TSs in the adopted fog nodes have been virtually disaggregated as shown in Figure 3. This improvement is enabled via better resource utilization achieved by server disaggregation. Disaggregation relaxes server-level resource locality constraints associated with TSs to enable per component resource utilization within each adopted computing location. This release compute resources in locations where there is enough network capacity to support the traffic demands of all workloads. Disaggregation enables about 33% and 13% reduction in the number of active components and servers respectively as shown in Figure 4 whilst ensuring no workload is blocked without increasing the capacity of network links. However, to ensure that optimal benefits of server disaggregation are achieved, the average hop count between active workload source and their corresponding computation location increased relative to the situation when TSs were deployed as shown in Table 3. This contributes to the marginal rise in total network traffic and the corresponding marginal rise in the total network power consumption shown in Figure 5 and Figure 6 respectively.

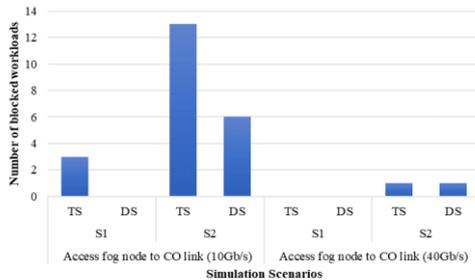

*Figure 3. Number of blocked workloads.*

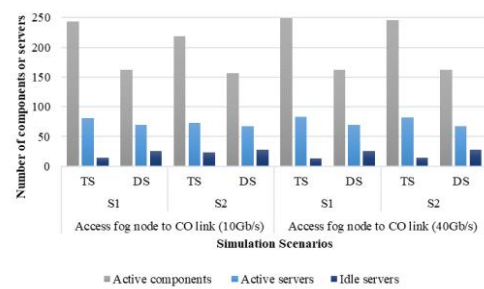

*Figure 4. Number of active/idle resources.*

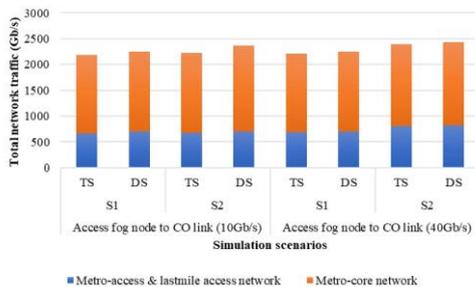

*Figure 5. Power consumption of metro and access networks.*

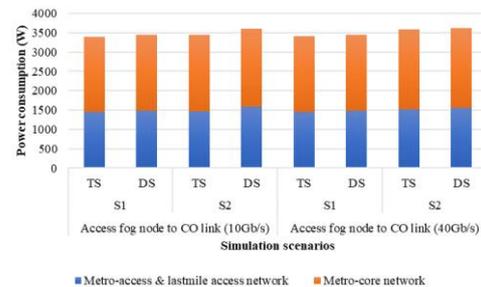

*Figure 6. Total traffic in metro and access networks.*

When the network traffic associated with interactive applications is scaled by a factor of two as in S2, the impact of network bottlenecks is further emphasised via the increase in the number of blocked workload when both TSs and DSs are deployed in the adopted fog nodes as shown in Figure 3. Of the 13 workloads blocked when TSs are used under S2, one workload is blocked because the access 10G EPON link between the user's residence and the

CO lacks enough capacity to support the network traffic associated with the workload and the regular network traffic. Concurrently, two workloads are blocked because the links between radio CSs and COs do not have enough capacity to support the traffic associated with these workloads and regular traffic concurrently. Finally, ten workloads are blocked because of insufficient network capacity to access stranded idle compute resources in CSs and BOs after COs compute resources have been optimally utilized. Of the idle servers, 74% are in BOs while 26% are in CSs. Disaggregation also reduces the number of blocked workloads by more than 50% because of improved efficiency in resource utilization which can be likened to increased cost and energy efficiencies. However, three workloads are still blocked because of network bottlenecks which prevent the use of idle servers in some BOs and CSs. Three workloads are also blocked because of insufficient capacity of 10G EPON links and metro Ethernet links to support traffic associated with workloads and regular network traffic concurrently as highlighted when TSs are employed in S2.

*Table 3. Average hop count between active workload source and processing location.*

|  | Access fog node to CO link (10Gb/s) | | | | Access fog node to CO link (40Gb/s) | | | |
|---|---|---|---|---|---|---|---|---|
| **Scenarios** | **S1** | | **S2** | | **S1** | | **S2** | |
| **Server Architecture** | TS | DS | TS | DS | TS | DS | TS | DS |
| **Average hop count** | 1.56 | 2 | 1.38 | 2 | 1.67 | 2 | 1.61 | 1.85 |

Figure 3 shows that scaling the capacity of links between COs and computing resources in BOs and CSs from 10 Gbps to 40 Gbps helps to alleviate workload blocking associated with network bottlenecks. Results under S1 show that all workloads are provisioned within the metro networks when either TSs or DSs are adopted for computation. However, as illustrated in Table 3, the average hop count between active workload source and processing location is higher when DSs are used. When TSs are used, workloads originating from CSs are often placed in their source CS to minimize overall network traffic and power consumption. This trend reduces when DSs are used because of the gains in compute resource utilization that remote placement of workloads can achieve. However, relative to the results obtained when TSs are used, overall network traffic in metro-access and last-mile access networks increased by 2.5% while the overall traffic in the metro-core network increased by 1.1% when DSs are used in S1 after link capacities have been scaled. This is responsible for the marginal rise in overall network power consumption as shown in Figure 5. However, 15.7% and 35% drop in the number of active servers and components are achieved respectively when DSs replace TSs in adopted fog computing locations i.e. COs, BOs and CSs. Figure 3 also shows that a workload is only blocked when either TSs or DSs are used after the link capacity between adopted fog nodes in CSs and BOs and COs have been scaled to 40 Gbps under S2 where workload traffic demands has doubled. This workload is blocked because the 10G EPON link connecting the source node of the workload to its adjacent COs lacks enough capacity to support the bandwidth demand of the workload and traditional network traffic concurrently. Hence, the network capacity of last-mile access links always forms part of the factor that may influence the feasibility of a FaaS system irrespective of the server architecture adopted in fog nodes. Furthermore, our result also shows that continuous rise in the number of workloads and their associated traffic can lead to bottlenecked links in the ring network topology adopted in the metro-core network.

## 4. CONCLUSIONS

In this paper, we evaluated the impact of using disaggregation as a tool for improved efficiency in a distributed FaaS system over telecom networks. Compared to the use of traditional servers, our results showed that disaggregation enables improved efficiency in fog node resource utilization at the cost of increased average hop count between workload source and processing location and marginal rise in overall network traffic and power consumption. The results also showed that network link capacity bottlenecks can prevent the use of compute resources in CSs and BOs despite server disaggregation especially when real-time workloads have high data rate. Scaling the capacity for links between COs and computing resources in CSs and BOs helps to mitigate this challenge. Future work will study the impact of real-time workloads with varying levels of latency sensitivity when disaggregated servers are deployed in a FaaS system.

**ACKNOWLEDGEMENTS**

The authors would like to acknowledge funding from the Engineering and Physical Sciences Research Council (EPSRC), INTERNET (EP/H040536/1) and STAR (EP/K016873/1) projects. The first author would like to acknowledge the support of the Petroleum Technology Trust Fund (PTDF), Nigeria, for the Scholarship awarded to fund his PhD. All data are provided in full in the results section of this paper.